\documentclass{ptephy_v1}

\preprintnumber{XXXX-XXXX} 

\begin{document}

\title{Precise measurement of the scintillation decay constant of ZnWO$_4$ crystal}


\author[1,*]{M. Shibata}
\author[1,2]{H. Sekiya}
\author[3]{K. Ichimura}
\affil[1]{Kamioka Observatory, Institute for Cosmic Ray Research, University of Tokyo, Hida, Gifu 506-1205, Japan}
\affil[2]{Kavli Institute for the Physics and Mathematics of the Universe (WPI), The University of Tokyo Institutes for Advanced Study, University of Tokyo, Kashiwa, Chiba 277-8583, Japan}
\affil[3]{Research Center for Neutrino Science, Tohoku University, Sendai 980-8578, Japan}
\affil[*]{\rm E-mail: mshibata@km.icrr.u-tokyo.ac.jp}

\begin{abstract}%
The scintillation decay time constant of a ZnWO$_4$ crystal irradiated with $\alpha$-particles from $^{241}$Am was precisely investigated, and was
found to depend on the incident direction of the $\alpha$-particles on the crystal.
The longest decay time constant ($24.3\pm0.6~\mu$s) was obtained on the surface perpendicular to the b-axis of the crystal (Surface B).
On Surfaces A and C, the decay constants were $20.0$ and $21.3\pm0.2~\mu$s, respectively. 
The scintillation yield of the ZnWO$_4$ was also anisotropic and depended on the incident direction of the heavy particles.
The maximum yield was achieved on surface B, suggesting a correlation between the light yield and scintillation decay time constant of ZnWO$_4$ crystals.

\end{abstract}

\subjectindex{H20, C40, C43}

\maketitle

\section{Introduction}
The existence of dark matter has been suggested by 
many observations of the universe at various scales. Among the dark matter candidates are weakly interacting massive particles (WIMPs), which are expected to be observed by elastic scattering with nuclei on the Earth~\cite{Drukier}.
Assuming that WIMPs in the galaxy follow the Maxwell distribution, the Earth is subjected to a WIMP wind caused by galactic rotation. Therefore, the most convincing signatures of WIMPs should appear in the directions of nuclear recoils~\cite{Spergel}, which can be monitored by detectors sensitive to these directions. 
Anisotropic scintillation crystals, in which the scintillation efficiency depends on the direction of heavily charged incident particles, have been investigated for this purpose~\cite{Belli1992,Spooner,Sekiya}.
With these crystals, the WIMPs' signal can be obtained by comparing
the visible energy spectra measured for different orientations related to the WIMP wind.
The anisotropic scintillation response of ZnWO$_4$ crystals was first reported in 2005~\cite{Danevich2005, Belli2011, Cappella2013} and has since been measured under irradiation with $\alpha$ and neutron sources ~\cite{Juan2020,Ichimura2020,Belli2020}. The radiopurity of ZnWO$_4$ is considered as an
advantage in dark matter detectors~\cite{Barabash2016,Belli2019}. 

ZnWO$_{4}$ is a colorless, transparent, monoclinic inorganic scintillation crystal. 
The basic properties of ZnWO$_4$ are tabulated in Table \ref{tb:property} and the length and angle parameters of the unit cell of ZnWO$_4$ are shown in Table \ref{tb:unit}.

\begin{table}
\begin{center}
  \caption{Properties of ${\rm ZnWO_4}$ scintillators}
  \label{tb:property}
  \centering
   \begin{tabular}{lc}
   \hline
    Molar mass & 313.22 ${\rm g/mol}$\\
    \hline
    Density & 7.87 ${\rm g/cm^3}$ \\
    \hline
    Melting point & 1200 \(^\circ\)C \\
    \hline 
    Mohs hardness & 4$\sim$4.5 \\
    \hline
    Reflective index & 2.1$\sim$2.2 \\
    \hline
    \end{tabular}
\end{center}
\end{table}

\begin{table}
\begin{center}
  \caption{Angles and lengths of the ZnWO$_4$ unit cell}
  \label{tb:unit}
  \centering
   \begin{tabular}{ccc}
    \hline
    $\alpha$[deg.] & $\beta$[deg.] & $\gamma$[deg.] \\
    90.0000 & 90.6210 & 90.0000 \\
    \hline \hline
    a[\AA] & b[\AA] & c[\AA] \\
    4.96060 & 5.718201& 4.92690 \\
    \hline
  \end{tabular} 
 \end{center}
\end{table}

ZnWO$_4$ scintillation crystals are also characterized by their
long decay time.
In the 1980's, the decay time constant of ZnWO$_4$ was reported as 20 $\mu$s under $\gamma$-ray
irradiation by Hol et al. ~\cite{Holl1988} and as 21.8 $\mu$s under X-ray irradiation by Grabmaier et al.~\cite{Grabmaier1984}.
In the 2000's, the decay time constant of ZnWO$_4$ was refined into three components: 25 $\mu$s, 7 $\mu$s, and 0.7 $\mu$s at 295 K under $\gamma$-ray irradiation from $^{137}$Cs~\cite{Nagornaya2009} and    
$25.7\pm 0.3 \mu$s, $5.6\pm 0.3 \mu$s, and $1.3 \pm 0.1 \mu$s at 295 K under $\alpha$-particle irradiation from $^{241}$Am~\cite{Kraus2005}. However, the crystal orientations were not identified.

As the scintillation yields of ZnWO$_4$ crystal under heavily charged particles depend on the incident direction, 
the scintillation decay time constant might similarly depend on the incident direction.
To test this idea, we systematically measured the scintillation decay time 
of a ZnWO$_4$ crystal in different orientations irradiated with $\alpha$-particles.

\section{Experimental Setup}
\begin{figure}[htbp]
  \begin{center}
   \includegraphics[width=120mm]{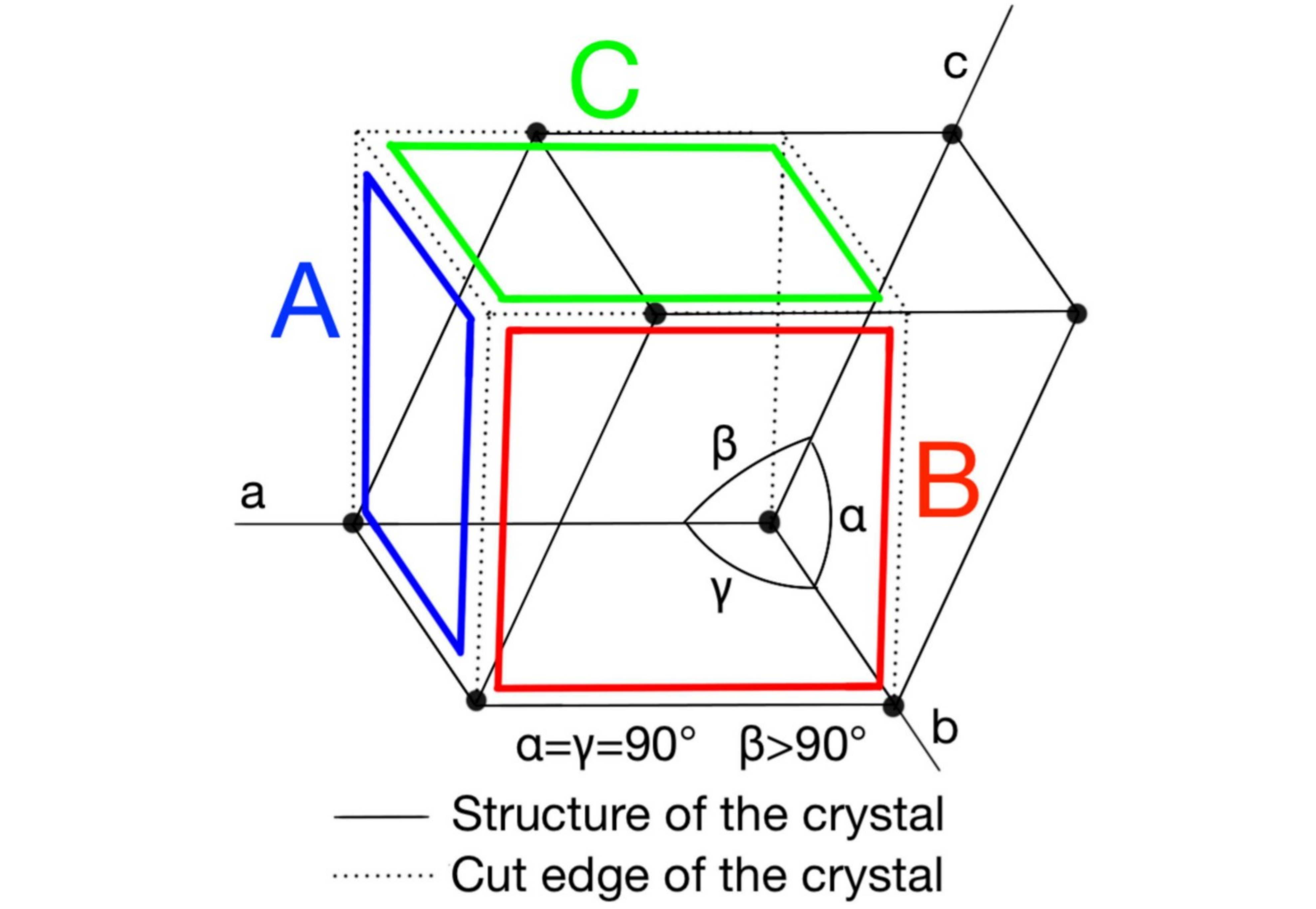}
  \end{center}
  \caption{Structure of a unit cell of the ZnWO$_4$ crystal (solid line) and the surfaces defined in the present experiment (dotted line). This graphic is taken from \cite{Juan2020}. Reprinted from our previous work.}
  \label{fig:surface_definition}
\end{figure}

The decay times were evaluated on a ZnWO$_4$ crystal with cubic dimensions (2 cm$\times$2 cm$\times$2 cm).
The crystal was produced at the Laboratory of Crystal Growth, Nikolaev Institute of Inorganic Chemistry in Russia and was cut by Crystal Manufacturing Lab. Ltd. (Russia) \cite{Ichimura2020}.
During cutting, the shape of the monoclinic unit cell was preserved as far as possible (see Fig.~\ref{fig:surface_definition}). The bottom surface was that of the unit cell and the surfaces were parallel to the unit cell. Surfaces A, B, and C were perpendicular to the a-, b-, and c-axes, respectively. The accuracy of the crystal cut was confirmed to be within 1$^{\circ}$~\cite{Juan2020}.

\begin{figure}[htbp]
  \begin{center}
   \includegraphics[width=50mm]{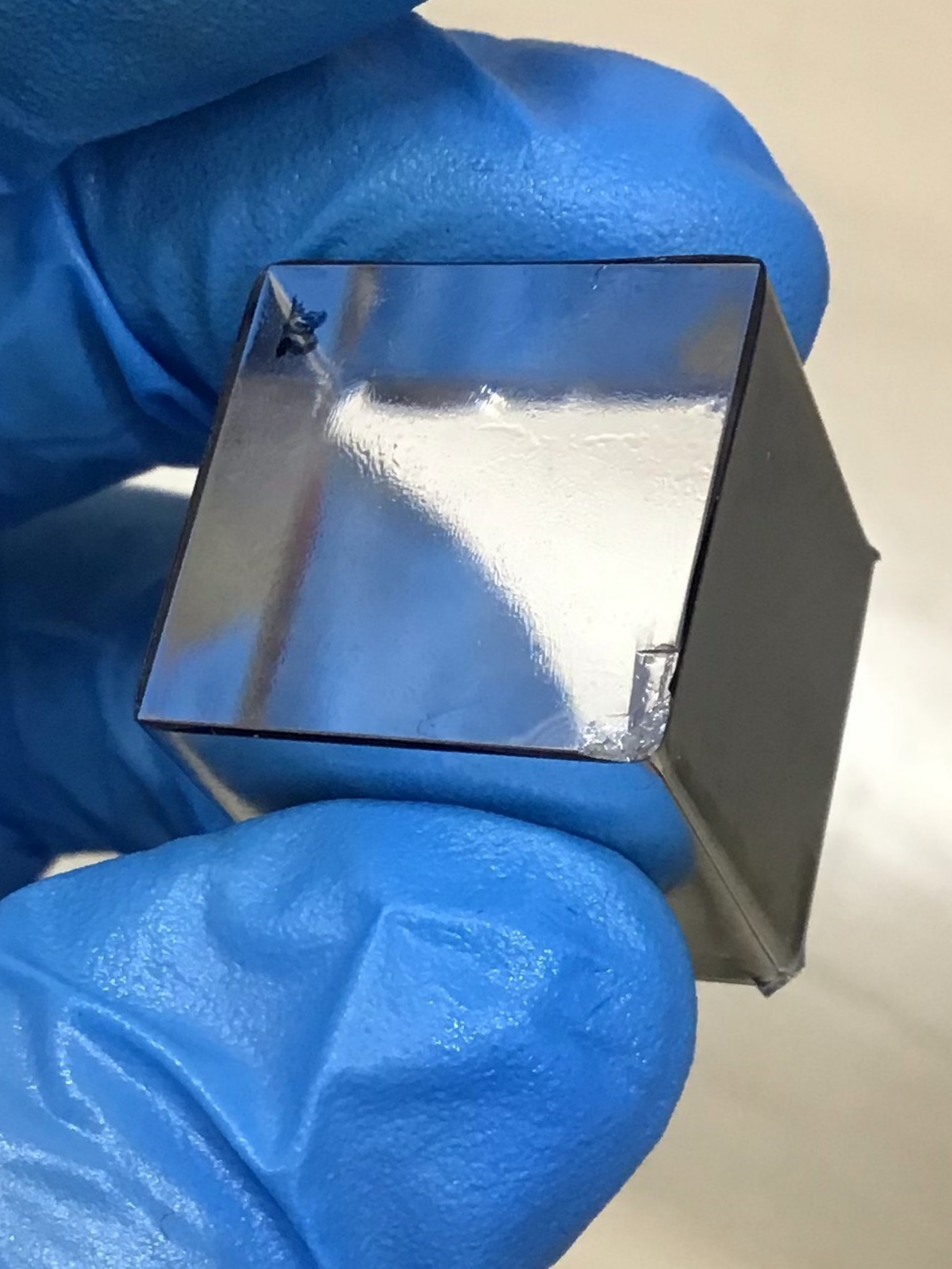}
   \includegraphics[width=60mm]{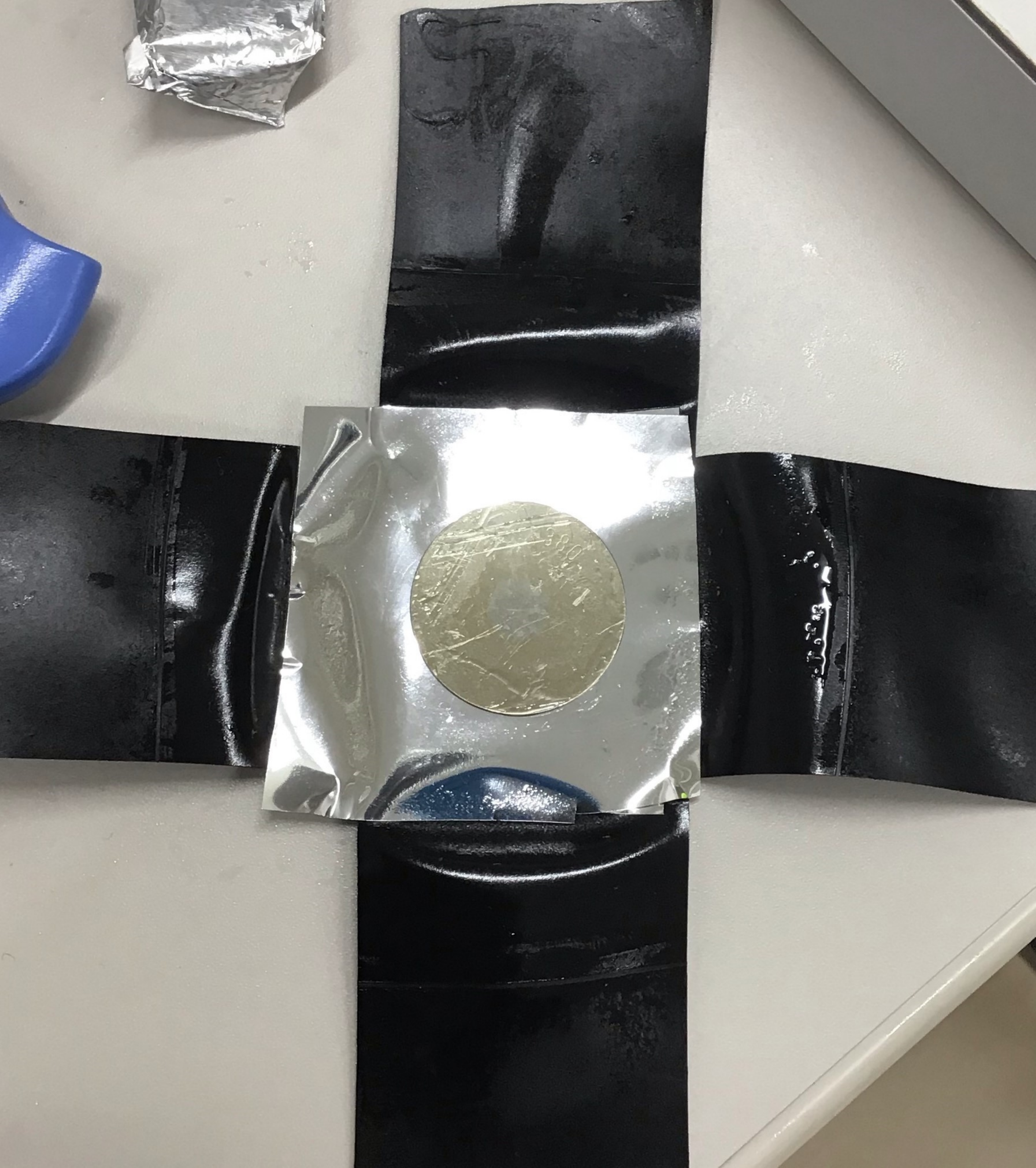}
  \end{center} 
  \caption{(left) ZnWO$_4$ crystal wrapped with the reflector; (right) $^{241}$Am source with the reflector.}
  \label{fig:detect_alpha1}
\end{figure}

To investigate scintillation decay in the ZnWO$_4$ crystal, the crystal was irradiated with a $\gamma$-ray source (708~kBq of $^{137}{\rm Cs}$) and an $\alpha$-particle source (2.9~kBq of $^{241}{\rm Am}$).
The side of the cubic crystal was wrapped with reflective sheets composed of 0.05~$\mu$m-aluminized 50~$\mu$m-thick polyethylene terephthalate film as shown in Fig.~\ref{fig:detect_alpha1}. For $\alpha$-particle irradiation, the $^{241}$Am source was placed in direct contact with the top of the crystal (surface A, B, or C) and was covered with the reflective sheet. The bottom of the crystal was attached to a 1''-$\phi$ PMT (Hamamatsu H6410) with optical grease (Adhesive Materials Group, V-788). The crystal was fixed with a mounter constructed from Styrofoam$^{\text{\textcircled{\scriptsize{\rm R}}}}$.
The $\gamma$-ray irradiation experiment was configured similarly, but the top of the crystal was directly covered with the reflective sheet and the $^{137}$Cs source was placed 20 cm from the crystal to prevent signal pile-up.

The PMT was operated at a gain of $1.25 \times 10^7$ and an applied voltage of $-$2200 V. Waveform signals from the anode were recorded with a digital storage oscilloscope at a sampling rate of 1GS/s (Tektronix TBS1064).
The measurements were conducted in the laboratory under temperature control at 21$^\circ$C.

\section{Results}
\begin{figure}[htbp]
  \begin{center}
   \includegraphics[width=120mm]{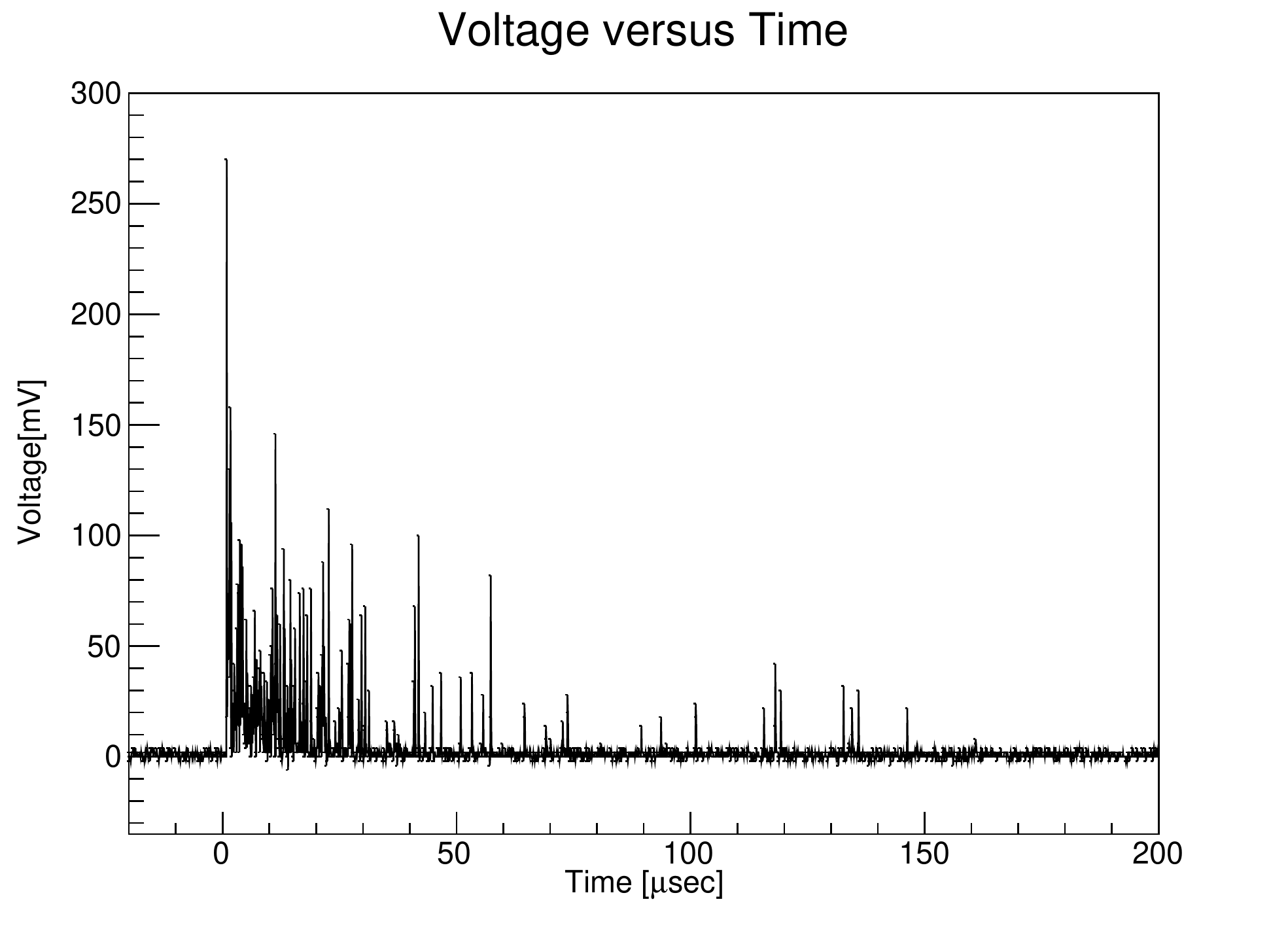}
  \end{center}
  \caption{Example of a scintillation decay waveform of ZnWO$_4$ irradiated with $\gamma$-rays from $^{137}$Cs, obtained by the oscilloscope. The PMT signal is discrete because the decay time is long.}
  \label{fig:decay_oscillo}
\end{figure}

A typical recorded waveform of the ZnWO$_4$ scintillator irradiated with $\gamma$-rays from $^{137}$Cs is shown in Fig.~\ref{fig:decay_oscillo}.
The waveform shows the typical structure of a train of photoelectrons distributed over tens of $\mu$s.
Therefore, to determine the precise decay time constants, the waveforms were statistically 
averaged over many waveforms. 
To this end, more than 6000 waveforms were recorded in each measurement. The total area of each waveform was calculated by integrating the data over the range 0 to 200 $\mu$s and an energy spectrum was obtained for each measurement.
Next, 
3000 waveforms of the 662~keV $\gamma$ events from $^{137}$Cs and 3000 waveforms of the 5.4~MeV $\alpha$ events from $^{241}$Am were selected and each waveform was normalized by its total area. The averaged waveforms were obtained by adding the 3000 normalized waveforms.
Finally the waveform was normalized again by the peak amplitude. 


The decay time constant was obtained by fitting the exponential functions of three components using the least-squares method.
The fitting equation was given by

\[y=A{\rm exp}\left\{-\frac{t-t_0}{\tau_1}\right\}+B{\rm exp}\left\{-\frac{t-t_0}{\tau_2}\right\}+C{\rm exp}\left\{-\frac{t-t_0}{\tau_3}\right\}+c\ .\]

\begin{figure}[htbp]
  \begin{center}
   \includegraphics[width=120mm]{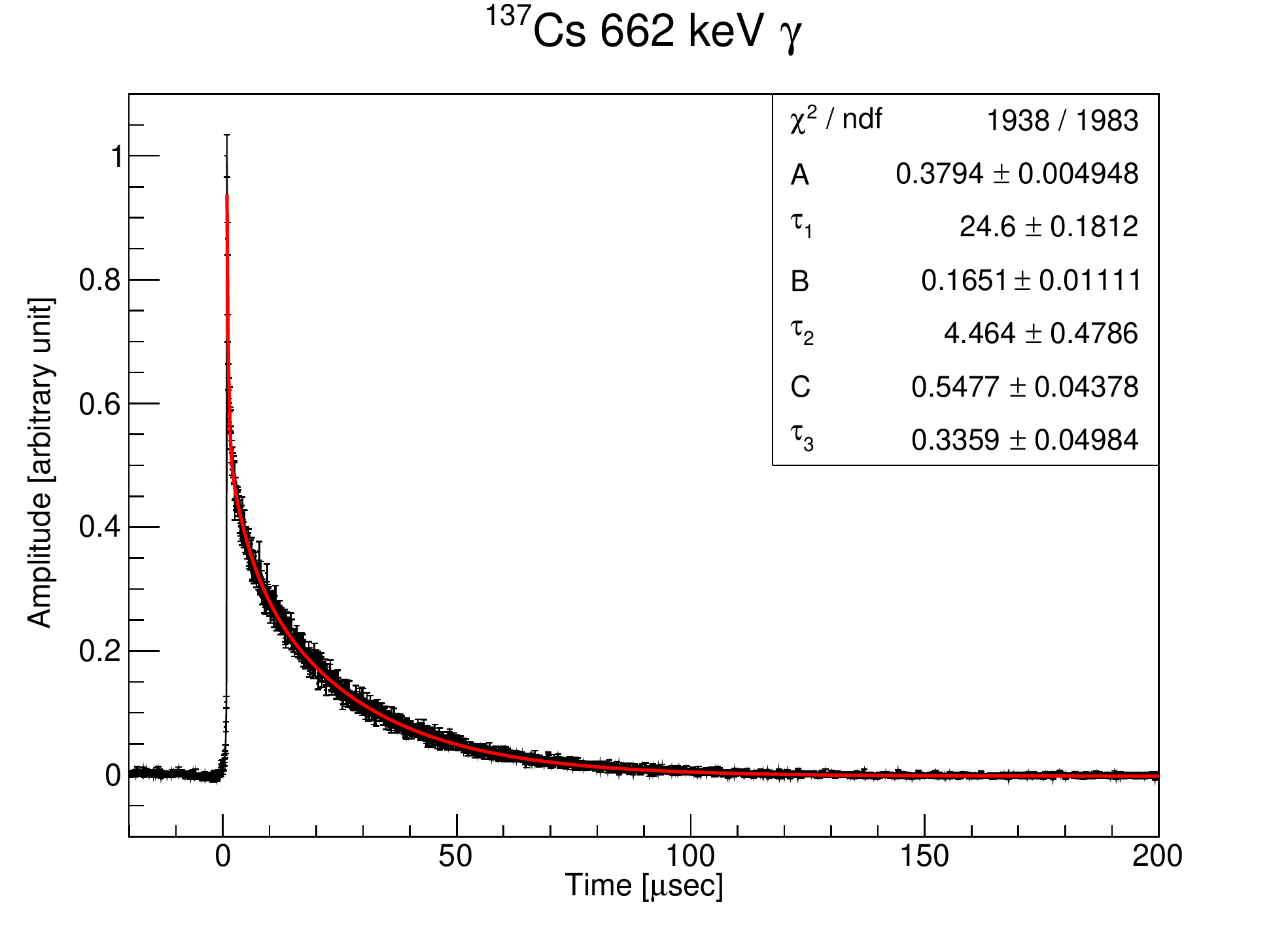}
  \end{center}
  \caption{Averaged waveform under $\gamma$-rays irradiation from $^{137}$Cs. The red line is the fitted function.}
  \label{fig:decay_G}
\end{figure}

Figure \ref{fig:decay_G} shows the obtained waveforms and the fitting results of $\gamma$-ray irradiation from $^{137}$Cs. The three components of the decay time were calculated as 24.6 $\pm$ 0.2 $\mu$s, 4.46 $\pm$ 0.48 $\mu$s and 0.34 $\pm$ 0.05 $\mu$s (see Table \ref{tb:alpha_timeconst}).

\begin{figure}[htbp]
  \begin{center}
   \includegraphics[width=120mm]{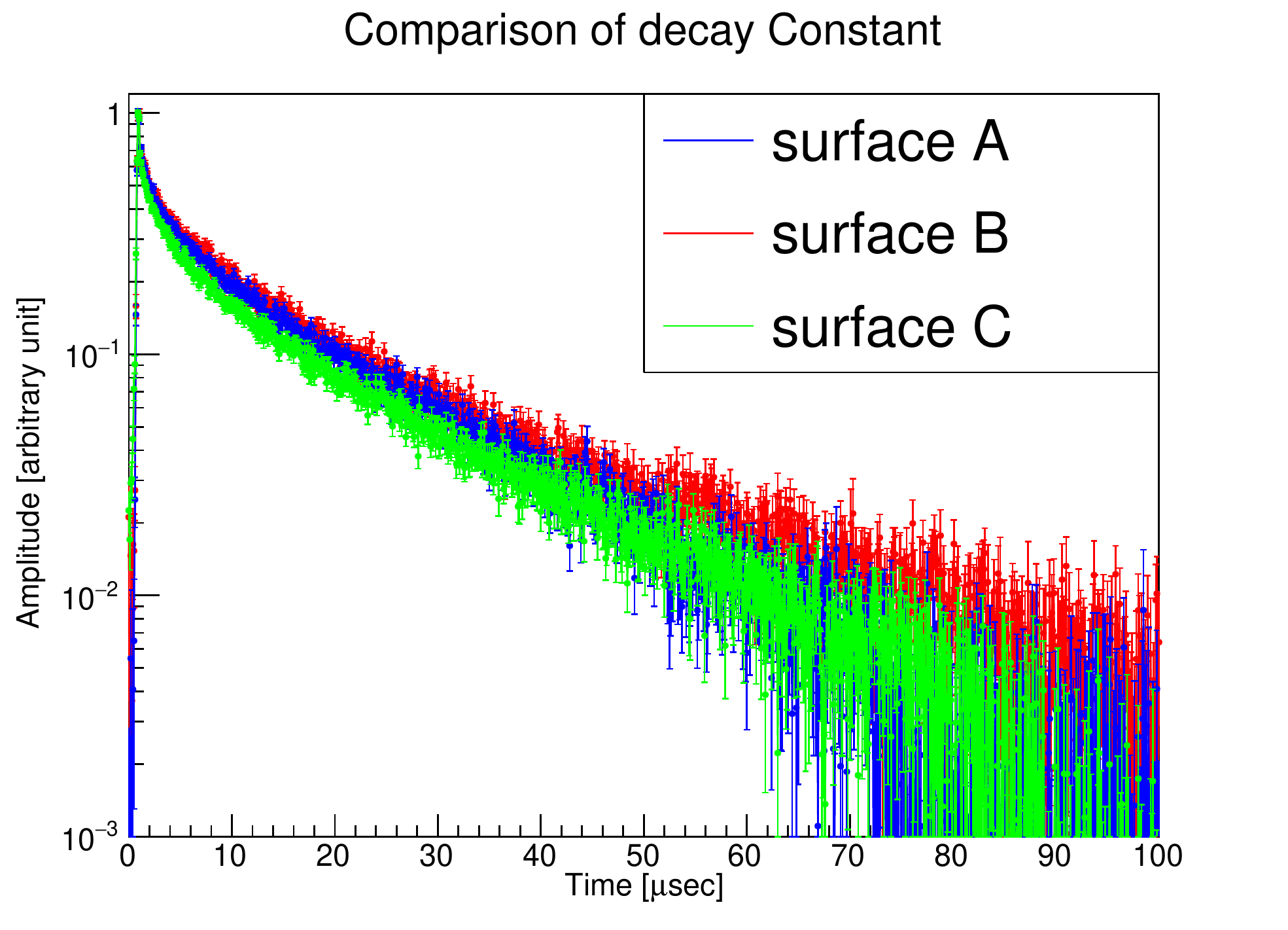}
  \end{center}
  \caption{Averaged waveforms of surfaces A (blue), B (red), and C (green) under $\alpha$-particle irradiation from $^{241}$Am to the surface A, B, and C. } 
  \label{fig:decay_comp}
\end{figure}

\begin{figure}[htbp]
  \begin{center}
   \includegraphics[width=120mm]{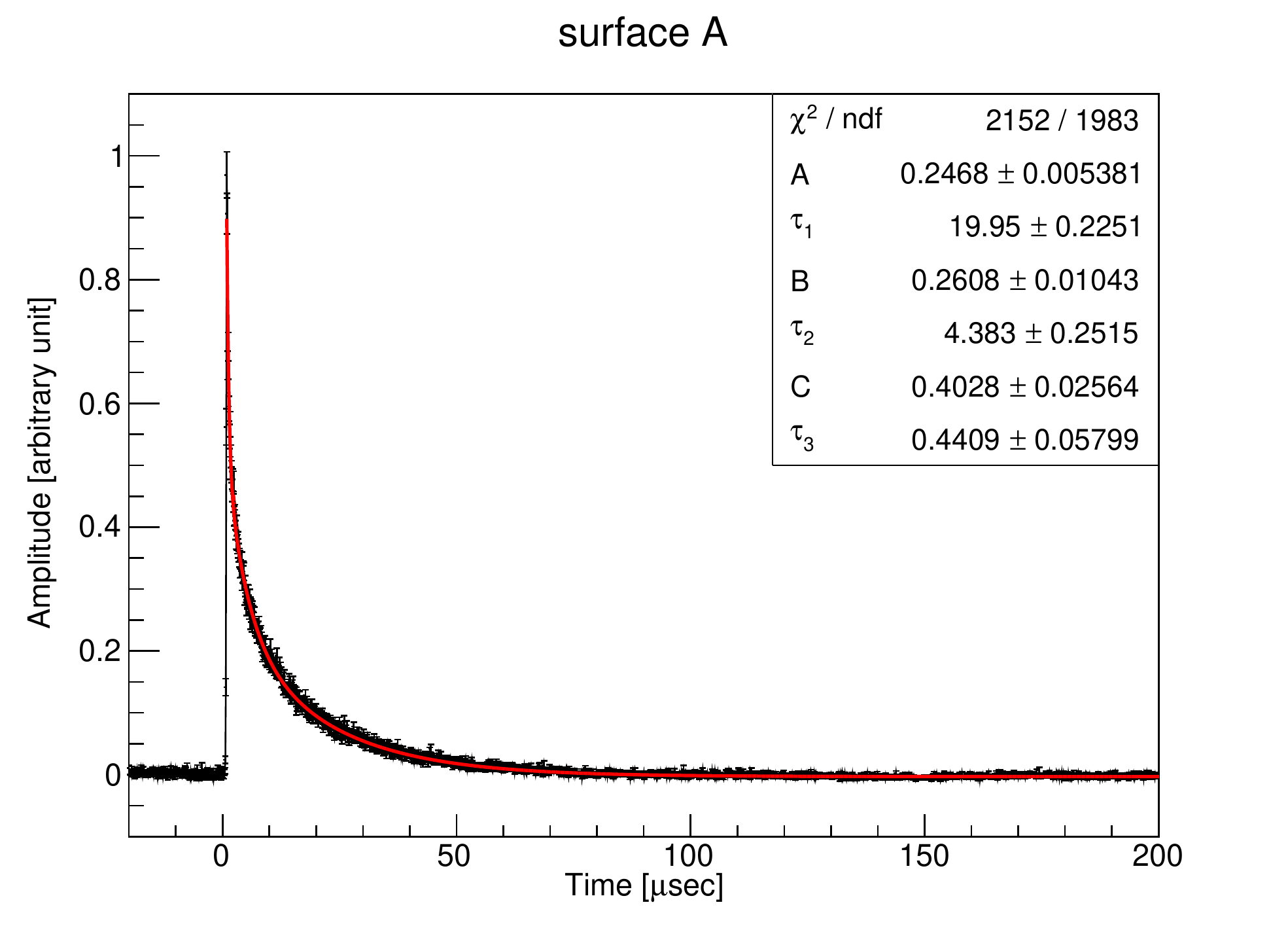}
  \end{center}
  \caption{Obtained averaged waveform of surface A irradiated with $\alpha$-particles 
  from $^{241}$Am. The red line is the fitted function.
  }
  \label{fig:decay_A}
\end{figure}

\begin{figure}[htbp]
  \begin{center}
   \includegraphics[width=120mm]{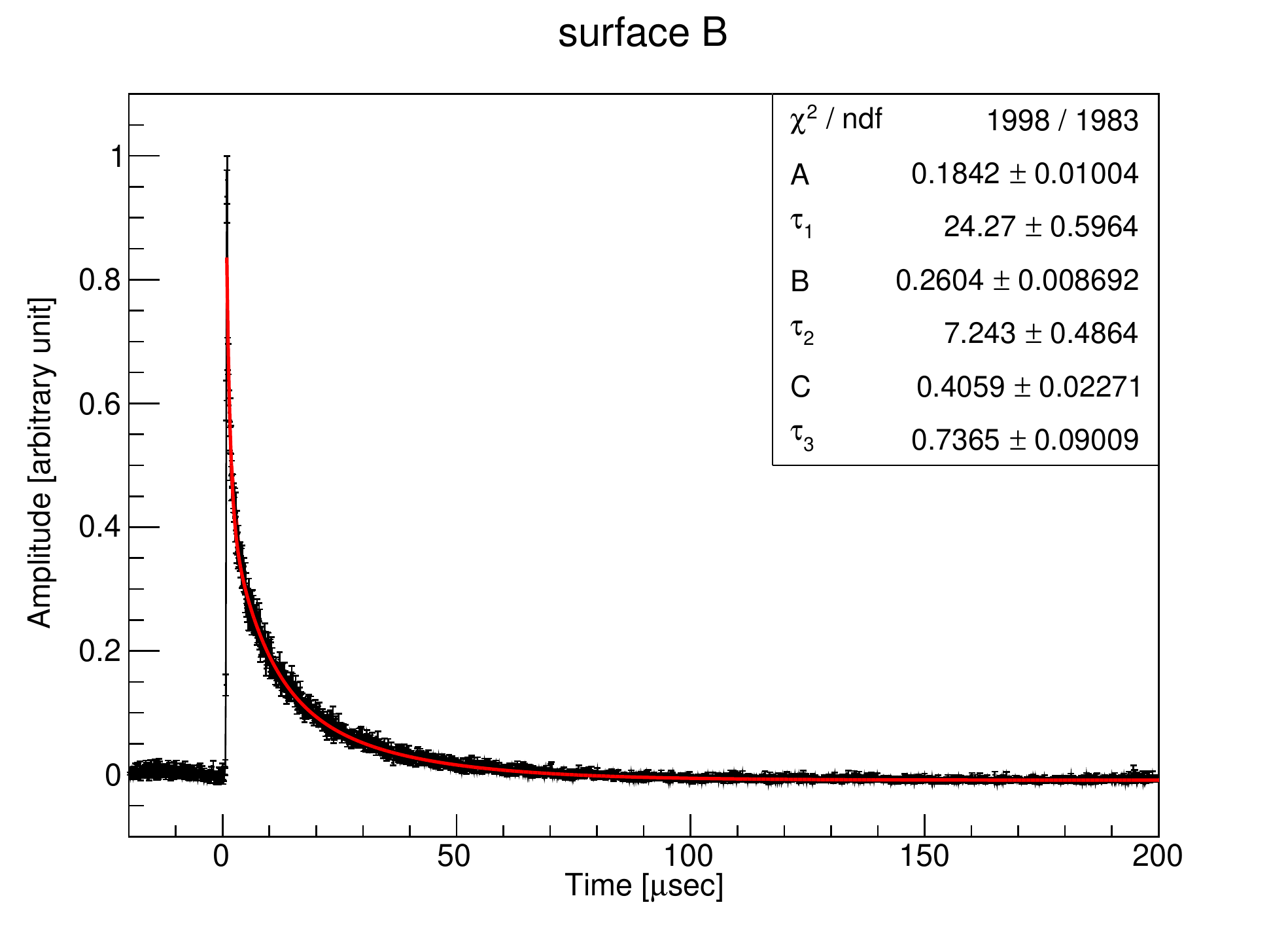}
  \end{center}
  \caption{Obtained averaged waveform of Surface B irradiated with $\alpha$-particles 
  from $^{241}$Am. The red line is the fitted function.}
  \label{fig:decay_B}
\end{figure}

\begin{figure}[htbp]
  \begin{center}
   \includegraphics[width=120mm]{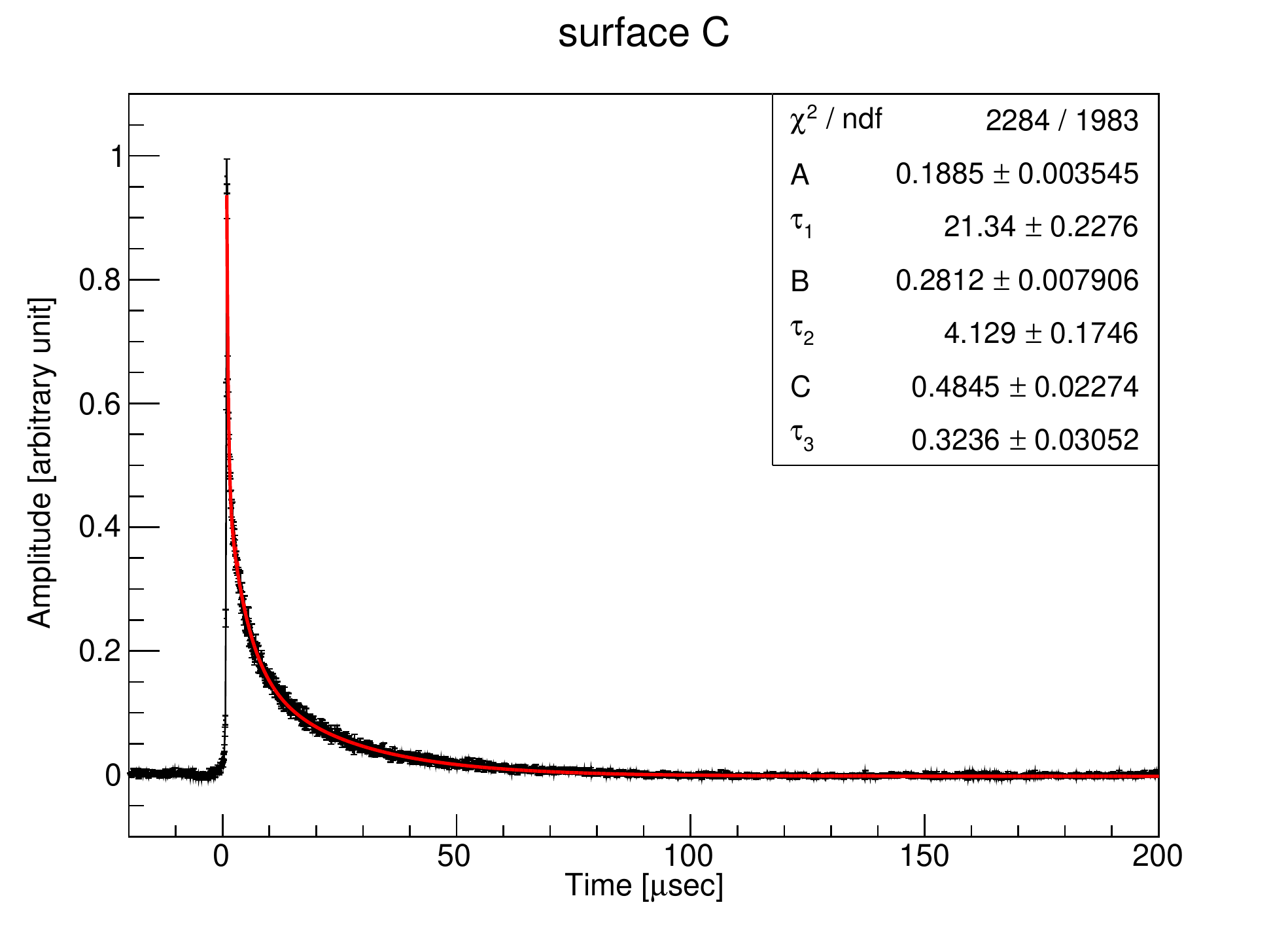}
  \end{center}
  \caption{Obtained averaged waveform on Surface C irradiated by $\alpha$-particles 
  from $^{241}$Am. The red line is the fitted function.}
  \label{fig:decay_C}
\end{figure}

The waveforms of surfaces A, B, and C irradiated with $\alpha$-particles from $^{241}$Am are shown in Fig. \ref{fig:decay_comp}, and their fitting results
are shown in Figs. \ref{fig:decay_A},\ref{fig:decay_B}, and \ref{fig:decay_C} respectively. The derived decay time constants are summarized in Table \ref{tb:alpha_timeconst}.
The longest components of the decay time constants at surfaces A, B, and C under the $\alpha$-particle irradiation were 20.0 $\pm$ 0.2 $\mu$s, 24.3 $\pm$ 0.6 $\mu$s, and 21.3 $\pm$ 0.2 $\mu$s, respectively.

\begin{table}[htbp]
 \caption{Scintillation decay constants of the ZnWO$_4$ crystal, obtained by fitting exponential functions to the three components}
 \label{tb:alpha_timeconst}
 \centering
  \begin{tabular}{|c|c|c|c|}
   \hline
   source & decay constant[$\mu$s] & $\chi^{2}$ / ndf & coefficient[$\times 10^{-1}$] \\
   \hline \hline
   $^{137}$Cs 
   & \begin{tabular}{c} 24.6$\pm$0.2\\ 4.46$\pm$ 0.48 \\ 0.34 $\pm$ 0.05\end{tabular} 
   & 1938/1983 
   & \begin{tabular}{c} $3.79\pm0.05$\\ $1.65\pm 0.11$ \\ $5.48\pm0.44$ \end{tabular} \\
   \hline
   $^{241}$Am A surface
   & \begin{tabular}{c} 20.0$\pm$0.2\\ 4.38$\pm$ 0.25 \\ 0.44 $\pm$ 0.06\end{tabular} 
   & 2152/1983 
   & \begin{tabular}{c} $2.47\pm0.05$\\ $2.61\pm 0.10$ \\ $4.03\pm0.26$ \end{tabular} \\
   \hline
   $^{241}$Am B surface 
   & \begin{tabular}{c}24.3$\pm$0.6 \\ 7.24$\pm$ 0.49 \\ 0.74$\pm$0.09 \end{tabular}
   & 1998/1983 
   & \begin{tabular}{c} $1.84\pm0.10$\\ $2.60\pm 0.09$ \\ $4.06\pm0.23$ \end{tabular} \\
   \hline
   $^{241}$Am C surface 
   & \begin{tabular}{c} 21.3$\pm$0.2\\ 4.13 $\pm$ 0.17 \\ 0.32 $\pm$ 0.03\end{tabular} 
   & 2284/1983 
   & \begin{tabular}{c} $1.89\pm0.04$\\ $2.81\pm 0.08$ \\ $4.85\pm0.23$ \end{tabular} \\
   \hline
  \end{tabular}
\end{table}

\section{Conclusion}
The anisotropic scintillation responses of a ZnWO$_4$ crystal to heavy particles ~\cite{Juan2020,Ichimura2020}
were investigated by irradiating each surface of the cubic crystal with $\alpha$-particles from $^{241}$Am and obtaining the scintillation decay time constants.
The main components of the time constants depended on the incident surface. The longest time constant was $24.3\pm0.6~\mu$s on surface B, where the light yield was also maximized. The other time constants were $20.0\pm0.2~\mu$s on surface A and  $21.3\pm0.2~\mu$s on surface C. When 662~keV $\gamma$-rays were irradiated on the crystal surface, the longest time constant was $24.6\pm0.1~\mu$s, close to that of surface B irradiated with $\alpha$-particles, and consistent with previously reported values 
~\cite{Nagornaya2009,Kraus2005}.
These measured time constants suggest that 
the anisotropy of the light yields originates from the scintillation decay process.
Using scintillation waveform information of ZnWO$_4$, in addition to the total light yields, may enhance the direction sensitivity to the WIMP wind.

\section*{Acknowledgment}

This work was supported by JSPS KAKENHI grant numbers 15K13478 and 17H02884 and by Grant for Basic Science Research Projects from The Sumitomo Foundation.


%

\vspace{0.2cm}
\noindent


\let\doi\relax


\end{document}